\documentclass[12pt]{article}
\begin{document}
\title{Quantum stability of defects for a Dirac field 
coupled to a scalar field in $2+1$ dimensions} 
\author{C.~D.~Fosco$^a$
  \\
  {\normalsize\it $^a$Centro At{\'o}mico Bariloche - Instituto Balseiro,}\\
  {\normalsize\it Comisi{\'o}n Nacional de Energ{\'\i}a At{\'o}mica}\\
  {\normalsize\it 8400 Bariloche, Argentina.}}  
\date{\today}
\maketitle
\begin{abstract}
\noindent We study the Euclidean effective action and the
full fermion propagator for a Dirac field in the presence of a scalar
field with a domain wall defect, in $2+1$ dimensions. We include
quantum effects due to both fermion and scalar field fluctuations, 
in a one-loop approximation. The results are interpreted in terms of
the quantum stability of the zero mode solution. We also study, for
this system, the induced `inertial' electric field for the fermions
on the defect, due to the quantum fluctuations of the scalar field.
\end{abstract}
\bigskip
\newpage
\section{Introduction}\label{intro}
The subject of fermionic fields in the background of defects presents
itself in many different areas of physics, from textures in superfluid
phases of $He_3$~\cite{he3} to cosmic strings~\cite{cosmic}. Those
defects are generally defined as some field configurations having a
nontrivial topological content, and one of the most important features
of these systems is the existence of localized fermionic zero modes.
The most interesting property of the zero modes, when transport
properties are considered, is the fact that they are perfect
conductors (namely, they have linear dispersion relations for small
values of the momenta).

The most widely known example of this phenomenon is, perhaps, the so
called Callan-Harvey mechanism~\cite{callan} by which a Fermi field
(either `fundamental' or arising as an effective variable in an
approximate description of a physical system) living in an
odd-dimensional spacetime, is coupled to a domain wall like defect.
This defect is defined as a mass term that changes sign on a
one-dimensional curve in the constant-time plane.  A general result
for this kind of system is that the fermionic field spectrum has,
besides the usual continuum part, an extra zero-mass bound-state
located precisely on the defect.  This zero mode corresponds then to a
one-dimensional fermion which, because of its linear dispersion
relation, is capable of carrying electric currents.

When the Fermi field is also minimally coupled to an Abelian gauge
field, the resulting Chern-Simons current, proportional to the
sign of the fermionic mass, has an apparent `anomaly' on the defect,
due to the vanishing of the mass. This `bulk' anomaly is, however,
exactly compensated by an equal and opposite (chiral) anomaly for the
fermionic current at the defect, which arises because the zero modes
are described by a $1+1$ dimensional {\em chiral\/} theory.  A similar
phenomenon is found at the edge of a droplet of a non-relativistic
electron gas in two spatial dimensions, in the presence of a magnetic
field, both in the regime of the fractional and integer quantum Hall
effects~\cite{frad}.  Here, the Chern-Simons (Hall) current is of course also
present, and its apparent `bulk' anomaly on the borders is canceled by
the gauge anomaly of the chiral edge states~\cite{wen,haldane}.
Indeed, this anomaly-matching mechanism is usually a guiding
principle for the construction of the models describing the matter
fields on the borders, since in two spacetime dimensions the form of
the anomaly is very closely constrained, though no completely
determined, by the matter content.

While for many different physical situations static defects are
sufficient to describe the relevant physics, in many others
(particularly in condensed matter physics) it is of current interest
to understand the possible effects due to the {\em dynamics\/} of the
defects. In particular, it is interesting to investigate the
interaction of the fermion zero modes with the fluctuating geometry
induced by the dynamics of the defect.  In~\cite{ffl} we generalized
the results of~\cite{fl}, which dealt with static defects in $2+1$
dimensions, to the case of an arbitrary, space {\em and time\/}
dependent defect, showing that an analog of the Callan and Harvey
mechanism~\cite{callan} still holds. Namely, there exists a chiral
zero mode, and it is localized, this time on the defect {\em
  worldsheet}.  Besides, an important new phenomenon is that (even
in the absence of any external gauge field) if the defect is
accelerated, there is an induced fermionic current on the defect. This
effect is caused by the gravitational chiral anomaly of the fermions
localized on the defect.  This current can be thought of as due to the
existence of an effective `inertial' electric field, which can be
entirely defined in terms of the geometry of the defect, and may also
be interpreted in terms of the gravitational anomaly for the fermions
living on the defect worldsheet~\cite{kks}.

In this paper, we shall consider a related situation: the defect will
be allowed to fluctuate, but the dynamics of these fluctuations will
be, in the present case, defined by a scalar field action.  To make
the system tractable, however, we shall deal with configurations where
the `topological charge' of the configurations is fixed, corresponding
to just {\em one\/} defect.  The excursions around these
configurations will be first taken into account by performing a loop
expansion around a static defect background.  Although the expansion
itself is a standard procedure, it will demand here the use of a
suitable set of coordinates, compatible with the (non translation
invariant) configuration of the system. After showing that the domain
wall plus zero mode configuration is a consistent classical
configuration, we shall show that they are stable under quantum
fluctuations of the fermionic and scalar fields.

When the fermion field is a Majorana rather than a Dirac spinor, the model
becomes supersymmetric. We show that a similar mechanism occurs in
such a case, although the model itself is less interesting from the
phenomenological point of view, since there is no `electric' current 
associated to the Majorana fermion.

The organization of this paper is as follows: in
section~\ref{sec:thesyst}, we define the system and explore some of
its fundamental properties. In section~\ref{sec:loop} we discuss some
non-trivial aspects of the loop expansion for the effective action,
applying this to the derivation of the full fermion propagator
including one-loop corrections in section~\ref{sec:ferp}. This result is
applied to the study of the stability of the zero mode solution under the
fluctuations. 

The effective inertial electric field, due to the accelerating defect
is discussed in section~\ref{sec:el}.

The supersymmetric version of the model is presented, for the sake of
completeness, in Appendix A. 
Section~\ref{sec:conc} presents our conclusions.

\section{The system}\label{sec:thesyst}
The system we shall deal with may be conveniently defined in terms of an 
Euclidean generating functional:
\begin{equation}
  \label{eq:defzgen}
{\mathcal Z}(j, {\bar\eta}, \eta) \;=\; \int {\mathcal D}\phi {\mathcal D}\psi \, 
{\mathcal D}{\bar\psi} \; \exp [ - S(\phi,{\bar\psi},\psi )\,+\,
\int d^3x (j\phi\,+\,{\bar\eta}\psi\,+\,{\bar\psi}\eta )]
\end{equation}
where the (Euclidean) action $S(\phi,{\bar\psi},\psi)$ can be written as:
\begin{equation}
  \label{eq:defs}
S(\phi,{\bar\psi},\psi)\;=\; S_B (\phi) \,+\, S_F ({\bar\psi},\psi,\phi)  
\end{equation}
with
\begin{equation}
  \label{eq:defsb}
S_B(\phi) \;=\; \int d^3x \,[\frac{1}{2} (\partial \cdot \phi)^2  \,+\, V(\phi) ]
\end{equation}
and
\begin{equation}
  \label{eq:defsf}
S_F({\bar\psi},\psi,\phi) \;=\; \int d^3x \,{\bar\psi}\,[\not\!\partial
\,+\, g \phi (x) ] \,\psi \;.
\end{equation} 
In Equation (\ref{eq:defsb}), $V(\phi)$ denotes a local potential,
which should of course be consistent with the existence of a domain
wall configuration for $\phi$, and whose form will be made more
explicit later on. The coupling constant $g$ controls the strength of
the interaction between the fermions and the scalar field.

The Euclidean Dirac matrices $\gamma_\mu$ have been chosen according
to the convention
\begin{equation}\label{dfgp}
\gamma_0\;=\;\sigma_3 \;\;\;\gamma_1\;=\;\sigma_1 
\;\;\;\gamma_2\;=\;\sigma_2\;,
\end{equation}
where the $\sigma_\alpha$'s denote the usual Pauli matrices.

In the absence of external sources, the classical equations of motion
for this system are:
\begin{equation}
  \label{eq:eqom1}
[\not\!\partial + g \phi(x) ] \psi \;=\;0 
\end{equation} 
and
\begin{equation}
  \label{eq:eqom2}
- \partial^2 \phi (x) \,+\, V'(\phi(x)) \,+\, g {\bar\psi}(x) \psi(x) \;=\; 0 \;,
\end{equation}
where $V' \equiv \frac{d V}{d\phi}$.  In order to have a fermionic
zero mode, one should demand the existence of a domain wall like
configuration for $\phi$. This, in turn, suggests that the potential
$V$ must be non quadratic, with the simplest choice being a quartic
function. Thus the domain wall would correspond to an ($x_1$) 
translation invariant kink solution.  This, of course, amounts 
to an infinite energy configuration in the infinite volume case. 
We have in mind, however, a finite system which can nevertheless 
be regarded  as infinite (for calculational simplicity) in most of the
objects we shall be interested in. 

Let us, for the sake of simplicity, consider a static configuration 
of this type; i.e., $\phi = \phi(x_2)$. Then (\ref{eq:eqom1}) and 
(\ref{eq:eqom2}) reduce to:
\begin{equation}
  \label{eq:redeq1}
[\gamma_0 \partial_0 + \gamma_1 \partial_1 + \gamma_2 \partial_2 + g \phi(x_2) ] \psi \;=\;0
\end{equation}
and
\begin{equation}
  \label{eq:redeq2}
- \frac{d^2\phi(x_2)}{dx_2^2} \,+\, V'(\phi(x_2)) \,+\,
 g {\bar\psi}(x_2) \psi(x_2) \;=\; 0 \;,
\end{equation}
respectively. To see that a zero mode is indeed a possible consistent
solution for this coupled system of equations, let us consider a field
configuration $\varphi_k(x_2)$ verifying
\begin{equation}
  \label{eq:redeq21}
- \frac{d^2\varphi_k(x_2)}{dx_2^2} \,+\, V'(\varphi_k(x_2)) \;=\; 0 \;,
\end{equation}
with a potential $V$, such that $\varphi_k(x_2)$ is kink-like, changing
sign at $x_2=0$, say (the `center' of the kink). Inserting this
configuration $\varphi_k$ in the place of $\phi$ in (\ref{eq:redeq1})
yields, by the standard Callan-Harvey mechanism, a chiral zero mode 
solution for the fermion field, and this in
turn implies the vanishing of ${\bar\psi}\psi$, since this term
involves the two chiralities.  This means that $\varphi_k(x_2)$ is also an
exact solution of the full system of coupled equations (\ref{eq:redeq1}) 
and (\ref{eq:redeq2}).

Let us now consider the more interesting issue of the quantum dynamics
corresponding to the above configurations. To that end, we first
introduce $W[j,{\bar\eta}, \eta]$, the generating functional of
connected Green's functions, which is defined by
\begin{equation}
  \label{eq:defw}
W[j,{\bar\eta},\eta] \;=\; \ln {\mathcal Z}[j,{\bar\eta},\eta]
\end{equation}
and the effective action $\Gamma$, the Legendre transform of $W$:
\begin{equation}
  \label{eq:defga}
\Gamma[\varphi,{\bar\chi},\chi]\;+\;W[j,{\bar\eta},\eta]\;=\; \int d^3x [j(x) \varphi(x) \,+\, {\bar\eta}(x) \chi(x) 
\,+\, {\bar\chi}(x) \eta (x) ]
\end{equation}
where $\varphi$, ${\bar\chi}$ and $\chi$ are the `classical fields':
\begin{equation}
  \label{eq:defclf}
\varphi(x) \;=\; \frac{\delta W}{\delta j(x)} 
\;\;\;\;
\chi(x) \;=\; \frac{\delta W}{\delta {\bar\eta}(x)} 
\;\;\;\;
{\bar\chi}(x) \;=\; - \frac{\delta W}{\delta\eta(x)} 
\end{equation}
and we adopted the convention that functional derivatives of Grassmann
objects always act {\em from the left of the corresponding object}.

\section{Loop expansion}\label{sec:loop}
A standard one-loop calculation of the effective action $\Gamma$
yields~\cite{zinn}
\begin{equation}
  \label{eq:onel}
\Gamma(\varphi,{\bar\chi},\chi) \;=\; S(\varphi,{\bar\chi},\chi)
\,+\, \Gamma_1(\varphi,{\bar\chi},\chi) \,+\,\ldots
\end{equation}
where $S$ is the classical action and $\Gamma_1$ denotes the
order-$\hbar$ correction, given explicitly by
\begin{equation}
  \label{eq:defg1}
\Gamma_1(\varphi,{\bar\chi},\chi)\;=\; \Gamma_1^a (\varphi) \,+\, \Gamma_1^b(\varphi,{\bar\chi},\chi)
\end{equation} 
where:
\begin{equation}
  \label{eq:defg1a}
\Gamma_1^a \;=\; -{\rm Tr}\ln{\mathcal D}
\end{equation}
and
\begin{equation}
  \label{eq:defg1b}
\Gamma_1^b\;=\;\frac{1}{2} {\rm Tr} \ln 
\left[ \frac{\delta^2 S_B(\varphi)}{\delta\phi(x_1)
\delta\phi(x_2)} \,-\,2 \, g^2 \, {\bar\chi}(x_1) {\mathcal D}^{-1}(x_1,x_2) 
\chi(x_2) \right] \;,
\end{equation}
with the definition:
\begin{equation}
  \label{eq:defcald1}
{\mathcal D}\;=\; \not \! \partial + g \varphi \;\;.
\end{equation}
The two terms contributing to $\Gamma_1$ have a different physical
interpretation: while $\Gamma_1^a$ corresponds to the energy of the
`distorted' Dirac vacuum for the fermionic field in the presence of
the scalar field $\varphi$, $\Gamma_1^b$ takes into account the bosonic
fluctuations around the classical kink background.  We note that,
besides the standard fluctuation operator
$\frac{\delta^2S_B}{\delta\phi(x_1)\delta\phi(x_2)}$ corresponding to a self-interacting
scalar field, there is also a correction due to the fermion-boson
interaction.

As we have already pointed out in the introduction, the field $\varphi$ will
be assumed to fluctuate around a background defect.  Rather than
considering arbitrary configurations for that defect, we shall
concentrate on those corresponding to a sufficiently smooth, space
dependent step-like defect.  Moreover, the choice of the configuration
will also be dictated by the practical requirement of making the
calculations simpler. Our first simplifying assumption is to consider
a rectilinear defect, with $\varphi$ being a function of only one of the
spatial coordinates, say $x_2$, and changing sign along the straight
line $x_2=0$. This will introduce in (\ref{eq:defg1}) an explicit
dependence on $x_2$ in both terms. This dependence is obviously
carried by the operator ${\mathcal D}$ defined in (\ref{eq:defcald1})
but also an $x_2$ dependence will be induced in the double functional
derivative of $S_B$ appearing in $\Gamma_1^b$, since it has to be evaluated
on the classical configuration for $\varphi$.  To simplify the task, taking
advantage of the translation invariance on the $x_0x_1$ plane, one
should be able to disentangle the dynamics of the fermions in two
pieces: one depending only on $x_2$, and the other having support
exclusively in the plane $x_2=0$. To this end, we note that the
operator ${\mathcal D}$ may be rewritten as:
\begin{equation}
  \label{eq:defcald2}
{\mathcal D} \;=\; (a+\not\!{\hat \partial}) {\mathcal P}_L  
+ (a^\dagger+\not\!{\hat \partial}) {\mathcal P}_R
\end{equation}
where $a, a^\dagger$ are operators acting on functions of $x_2$,
\begin{equation}
  \label{eq:defaad}
a\;=\;\partial_2 + g \varphi(x_2) \;\;,\;\; 
a^\dagger\;=\;-\partial_2 + g \varphi(x_2) \;,
\end{equation}
${\mathcal P}_{L,R}$ are projectors along the eigenspaces of the
`chirality' matrix $\gamma_2$:
\begin{equation}
  \label{eq:defplr}
{\mathcal P}_L \;=\; \frac{1}{2} (1 + \gamma_2) \;\;\;\;
{\mathcal P}_R \;=\; \frac{1}{2} (1 - \gamma_2) \;,
\end{equation}
and $\not \! {\hat \partial}$ is the two-dimensional Euclidean Dirac
operator corresponding to the two coordinates $x_0$ and $x_1$, which
we denote collectively by ${\hat x}$, namely
\begin{equation}
  \label{eq:defsmalld}
\not \! {\hat \partial}\;=\; {\hat \gamma} \cdot {\hat \partial} \,=\, \gamma_\alpha \partial_\alpha \;\;\;, 
\;\;\; \alpha = 0, 1\,.
\end{equation}

Expression (\ref{eq:defcald2}) suggests that one could get rid of the
$x_2$ dependence of the fields by a suitable expansion in the modes of
some operator, thus obtaining a `dimensional reduction' from the three
dimensional spacetime to the two dimensional one described by ${\hat
  x}$. As ${\mathcal D}$ itself is not Hermitian, we may use instead
any of the positive Hermitian operators ${\mathcal H}$ or
${\widetilde{\mathcal H}}$:
\begin{equation}
  \label{eq:defhht}
{\mathcal H} \;=\; {\mathcal D}^\dagger {\mathcal D} \;\;,\;\;
{\widetilde{\mathcal H}} \;=\; {\mathcal D} {\mathcal D}^\dagger 
\end{equation}
to define the modes.

Using the expression (\ref{eq:defcald2}) for ${\mathcal D}$, leads to
the explicit form for ${\mathcal H}$ and ${\widetilde{\mathcal H}}$:
\begin{equation}
  \label{eq:defh1}
{\mathcal H} \;=\; - \partial^2 + g^2 \varphi^2 - g \gamma_2 \partial_2 
\varphi \;\;,\;\; {\tilde{\mathcal H}} \;=\; - \partial^2 + g^2 \varphi^2 
+ g \gamma_2 \partial_2 \varphi \;.
\end{equation}
These expressions enable us to write the contribution $\Gamma_1^a$ to
the effective action either in terms of ${\mathcal H}$ or 
${\widetilde{\mathcal H}}$:
\begin{equation}
  \label{eq:gma1}
\Gamma_1^a(\varphi) \;=\; - \frac{1}{2}{\rm Tr}\ln{\mathcal H}\;=\; 
- \frac{1}{2}{\rm Tr}\ln {\widetilde{\mathcal H}} \;.
\end{equation}
Taking the trace over Dirac indices, and using any of the two
previous representations, we end up with:
\begin{equation}
  \label{eq:gma11}
\Gamma_1^a(\varphi) \;=\; -\frac{1}{2} {\rm Tr} \ln ( -\partial^2 
+ g^2 \varphi^2 + g \partial_2\varphi ) -\frac{1}{2} {\rm Tr} \ln ( 
-\partial^2 + g^2 \varphi^2 - g \partial_2\varphi )
\end{equation}
where the trace is now meant only on functional space.

On the other hand, in $\Gamma_1^b$ we may factorize a bosonic fluctuation 
determinant, so that
$$
\Gamma_1^b \,=\, \frac{1}{2} {\rm Tr} \ln \left[ \frac{\delta^2
    S_B}{\delta\varphi(x_1)\delta\varphi(x_2)}\right]
$$
\begin{equation}
  \label{eq:gmb1}
+\, \frac{1}{2} {\rm Tr} \ln \left[ \delta(x_1-x_2) 
- 2 g^2 \int_y \Delta_\varphi(x_1,y) {\bar \chi}(y)
{\mathcal D}^{-1}(y,x_2) \chi(x_2) \right] \;,
\end{equation}
where:
\begin{equation}\label{eq:delphi}
\int_y \Delta_\varphi(x_1,y) \frac{\delta^2 S_B}{\delta\varphi(y)\delta\varphi(x_2)}=\int_y \frac{\delta^2 S_B}{\delta\varphi(x_1)\delta\varphi(y)}
\Delta_\varphi(y,x_2)=\delta(x_1-x_2) \;.
\end{equation}

With equation (\ref{eq:gmb1}) in mind, we note that the one-loop
effective action, $\Gamma_1$, may also be written as:
\begin{equation}
  \label{eq:newg1}
\Gamma_1 \;=\; \Gamma_1^{(0)} \,+\, \Gamma_1^{(I)} \;,
\end{equation} 
where $\Gamma_1^{(0)}$ represents the contribution that would
correspond to a system consisting of a free fermion in a non-trivial
$\varphi$ background plus the vacuum energy due to the
self-interacting scalar field $\varphi$, with no term taking into
account the interaction energy between scalar and fermion fields:
$$\Gamma_1^{(0)} \;=\; -\frac{1}{2}\, {\rm Tr}\ln\left[
  -\hat{\partial}^2 - \partial_2^2 + g^2 \varphi^2(x_2) - g \partial_2
  \varphi(x_2) \right] $$
$$-\frac{1}{2}\, {\rm Tr}\ln\left[-\hat{\partial}^2 - \partial_2^2 +
  g^2 \varphi^2(x_2) + g \partial_2 \varphi(x_2) \right] $$
\begin{equation}
+\;\frac{1}{2}\, {\rm Tr}\ln\left[-\hat{\partial}^2 - \partial_2^2 
+ V''(\varphi(x_2)) \right] \;.
\end{equation}
The $\Gamma_1^{(I)}$ term, instead, is a measure of the interaction
energy between bosons and fermions, vanishing when $g \to 0$:
\begin{equation}
  \label{eq:gmi}
\Gamma_1^{(I)}\;=\; \frac{1}{2} {\rm Tr} \ln \left[ \delta(x_1-x_2) 
- 2 g^2 \int_y \Delta_\varphi(x_1,y) {\bar \chi}(y)
{\mathcal D}^{-1}(y,x_2) \chi(x_2) \right] \;.
\end{equation}

In what follows, we shall discuss the contributions $\Gamma_1^{(0)}$
and $\Gamma_1^{(I)}$ separately.
We first make the observation that the `vacuum' term $\Gamma_1^{(0)}$
may be conveniently put in the following form:
\begin{equation}
\Gamma_1^{(0)} \,=\, -\frac{1}{2}{\rm Tr}\ln\left[-\hat{\partial}^2 + h_f \right]
-\frac{1}{2}{\rm Tr}\ln\left[-\hat{\partial}^2+{\tilde h}_f \right]
+\frac{1}{2}{\rm Tr}\ln\left[-\hat{\partial}^2 + h_\varphi \right] \;,
\end{equation}
where the dependence on the domain wall profile $\varphi (x_2)$ and on
the potential $V(\varphi(x_2))$ is encoded in the operators $h_f$,
${\tilde h}_f$ and $h_\varphi$, which are defined by
\begin{equation}
h_f \,=\, a^\dagger a \;\;,\;\; {\tilde h}_f \,=\,aa^\dagger \;\;,\;\;
h_\varphi \,=\, - \partial_2^2 + V''(\varphi(x_2)) \;.
\end{equation}
These operators play the role of  one dimensional quantum mechanical 
`Hamiltonians', whose eigenvalues appear as parameters in
the corresponding functional determinants. The eigenvalues and 
eigenvectors of these operators are defined by the equations
$$
h_f \psi_n(x_2) \;=\; \lambda_n^2 \psi_n(x_2) \;\;\;\;\;\;\; {\tilde h}_f {\tilde \psi}_n(x_2) \;=\;
\lambda_n^2 {\tilde \psi}_n(x_2) 
$$
\begin{equation}
h_\varphi \xi_n(x_2) \;=\; \mu_n^2  \xi_n(x_2) \;,
\end{equation}
where all the eigenvalues are written as squares (of real numbers) to
emphasize the fact that the corresponding operators are positive.

We now start to consider some simplifying assumptions regarding the
shape of the defect. Noting that
\begin{equation}
  \label{eq:gma12}
\Gamma_1^a(\varphi) \;=\; -\frac{1}{2} {\rm Tr} \ln ( -{\hat \partial}^2 + h )
-\frac{1}{2} {\rm Tr} \ln ( -{\hat \partial}^2 + {\tilde h} )
\end{equation}
where $h=a^\dagger a$, ${\tilde h}=a a^\dagger$ and $a=\partial_2 + g \varphi$, we assume that
only $h$ has a zero mode $|0\rangle$, which of course verifies the first
order equation $a |0\rangle = 0$. The rest of the $h$ and ${\tilde h}$
spectra coincide, and we demand it to be a continuum, separated by a
finite gap $\kappa^2>0$ from the zero mode.  The resulting equations
uniquely fix the form of $\varphi$ to be:
\begin{equation}
  \label{eq:tanh}
\varphi(x_2) \;=\; \frac{\kappa}{g} \, \tanh (\kappa x_2) \;,
\end{equation}    
where the constant $\kappa$ remains arbitrary. The first order equation 
for the zero mode can be integrated by quadratures, yielding:
\begin{equation}
  \label{eq:zm}
\psi_0(x_2)\;=\; \langle x_2|0\rangle \;=\; \sqrt{\frac{\kappa}{2}} \, [\cosh(\kappa x_2)]^{-1} \;, 
\end{equation}
which has been normalized to $\langle0|0\rangle=1$.
For the previous $\varphi$ field profile, we see that the explicit forms 
of the operators $h_f$ and ${\tilde h}_f$ become
\begin{equation}
  \label{eq:hfexp} 
h_f\,=\, -\partial_2^2 + \kappa^2 [ 2 \tanh^2(\kappa x_2) - 1]
\;\;\;\;\;\;
{\tilde h}_f\;=\; -\partial_2^2 + \kappa^2 \;.
\end{equation}
Although it is not immediately evident from (\ref{eq:hfexp}), the 
spectra of $h_f$ and ${\tilde h}_f$ coincide, except for the zero in $h_f$.
Indeed, any non-zero mode of ${\tilde h}_f$ may be used to generate an 
eigenvector of $h_f$ (with identical eigenvalue), by application of 
$a^\dagger$, and reciprocally, applying $a$ to a non-zero mode of $h_f$
produces an eigenvector of ${\tilde h}_f$. On the other hand, by 
assumption, only $h_f$ has a zero mode.  
Taking this into account, we see that $\Gamma_1^a$ may be written as
\begin{equation}
\Gamma_1^a \;=\; -{\rm Tr}_{2+1}\ln(-{\hat \partial}^2 - \partial_2^2 + \kappa^2) 
- \frac{1}{2} {\rm Tr}_{1+1}\ln(-{\hat\partial}^2) 
\end{equation}
where the dimension of the spacetime where the functional trace is defined has
been explicitly shown. Of course, we may also go back to a `spinorial' 
representation:
\begin{equation}
\Gamma_1^a \;=\; -{\rm Tr}_{2+1}\ln(\not\!\partial + \kappa) \,
-\,{\rm Tr}_{1+1}\ln(\not\!{\hat \partial}{\mathcal P}_L) \;, 
\end{equation}
where the zero mode contribution has been singled out, and the trace 
affects spinorial indices also.    

Fixing (\ref{eq:tanh}) to be the form of $\varphi$, we now need to
make sure that it is a minimum of $S_B$. This may be achieved by a 
judicious choice of the potential $V$. 
As already advanced, the simplest choice is to use a quartic potential; 
indeed, assuming the form of $V(\varphi)$ to be:
\begin{equation}
  \label{eq:vform}
V(\varphi) \;=\; \frac{\lambda}{2} \, (\varphi^2 - \mu^2)^2 \;,
\end{equation}
then there is a classical kink solution
\begin{equation}
  \label{eq:ksol}
\varphi(x_2) \;=\; \mu \, \tanh (\sqrt{\lambda}\, \mu x_2) \;.
\end{equation}
In order to match the configuration (\ref{eq:tanh}) (something we
require in order to simplify the spectra of $h$ and ${\tilde h}$), 
we should demand:
\begin{equation}
  \label{eq:fparm}
\mu \;=\; \frac{\kappa}{g} \;\;,\;\; \sqrt{\lambda} \mu \;=\; \kappa 
\end{equation}
so that the scalar potential is:
\begin{equation}
  \label{eq:vform1}
V(\varphi) \;=\; \frac{g^2}{2} \, \left(\varphi^2 - (\frac{\kappa}{g})^2\right)^2 \;.
\end{equation}

With $V$ as given by (\ref{eq:vform1}), we see that $h_\varphi$ is
\begin{equation}
h_\varphi \;=\; -\partial_2^2 \,+\, 6 \kappa^2 \tanh^2(\kappa x_2) - 2 \kappa^2 \;.
\end{equation}
The spectrum of $h_\varphi$ can also be found  by algebraic means.
We note that $h_\varphi$ can be factorized similarly to $h_f$:
\begin{equation}\label{eq:hphif}
h_\varphi \,=\, b^\dagger b 
\end{equation}
where we introduced: 
\begin{equation}
b \,=\, \partial_2 + 2 \kappa \tanh (\kappa x_2) \;\;\;,\;\;\;
b \,=\, -\partial_2 + 2 \kappa \tanh (\kappa x_2) \;.
\end{equation}
This already shows that there will be a zero mode $\xi_0$ for $h_\varphi$,
which verifies $b \xi_0 =0$, and is given by:
\begin{equation}\label{eq:phiz}
\xi_0(x_2)\;=\; \sqrt{\frac{3 \kappa}{4}} \, [\cosh (\kappa x_2)]^{-2}\;,
\end{equation}
when normalized to $\langle\xi_0|\xi_0\rangle =1$. Regarding the spectrum above the zero mode, we may take advantage of the fact that the operator 
${\tilde h}_\varphi \equiv b b^\dagger$ verifies:
\begin{equation}
{\tilde h}_\varphi \;=\; h_f \,+\, 3 \kappa^2 \,
\end{equation}
to infer that $h_\varphi$ will also have another bound state (with 
eigenvalue $3 \kappa^2$), and then a continuum starting at $4 \kappa^2$.
The form of these extra states may of course be found by applying
the operator $b^\dagger$ to the corresponding eigenstates of $h_f$. 

Thus the spectrum of $h_\varphi$ contains  two discrete states of 
eigenvalues $0$ and $3 \kappa^2$, plus a continuum starting at
$4 \kappa^2$.
Using this information in combination with the already known form of
$\Gamma_1^a$, we see that
$$
\Gamma_1^{(0)}\;=\; - \,\frac{1}{2} {\rm Tr}_{1+1}\ln(-{\hat\partial}^2)
\,+\,\frac{1}{2} {\rm Tr}_{1+1}\ln(-{\hat\partial}^2)
\,+\,\frac{1}{2} {\rm Tr}_{1+1}\ln(-{\hat\partial}^2 + 3 \kappa^2)
$$
\begin{equation}
-{\rm Tr}_{2+1}\ln(-{\hat \partial}^2 - \partial_2^2 + \kappa^2) 
\,+\, \frac{1}{2}{\rm Tr}_{2+1}\ln(-{\hat \partial}^2 - \partial_2^2 + 4 \kappa^2) \;, 
\end{equation}
where all the terms traces affect only functional space.
Note that the first two terms, which correspond to the fermionic
and bosonic zero modes, cancel out exactly.

\section{The full fermion propagator}\label{sec:ferp}
The full fermion propagator ${\mathcal S}_F$ may be determined by
evaluating the inverse of the second functional derivative of the
(one-loop corrected) effective action, evaluated at zero fermionic
field, namely,
\begin{equation}
  \label{eq:fullprop}
{\mathcal S}_F \;=\; {\tilde{\mathcal D}}^{-1} 
\end{equation}
where
\begin{equation}
  \label{eq:fer2} 
{\tilde{\mathcal D}}(x,y)\;=\;-\frac{\delta^2\Gamma}{\delta{\bar\chi}(x)
\delta\chi(y)}|_{\chi={\bar\chi} =0}
\;=\;{\mathcal D}(x,y) \;+\; {\mathcal D}_1(x,y)
\end{equation}
with
\begin{equation}
{\mathcal D}_1(x,y)\;=\; g^2\,\Delta_\varphi (y,x)\, {\mathcal D}^{-1}(x,y) \;.         
\end{equation}
It is evident that ${\tilde{\mathcal D}}$ plays the role of a
`corrected Dirac operator', since it includes a one-loop contribution.
Contrary to what happens for the translation invariant case, we see
that the use of a standard Fourier representation for all the
coordinates does not simplify the treatment. Rather, a mixed expansion
using plane waves for the ${\hat x}$ coordinates and suitable
eigenfunctions for the dependence on $x_2$ is more adequate. Indeed,
in terms of these eigenfunctions and eigenvalues, we may formulate the
expansion for the propagator ${\mathcal D}^{-1}$ as follows
$$
\langle x|{\mathcal D}^{-1}|y\rangle\;=\; \int \frac{d^2{\hat
    p}}{(2\pi)^2} \,e^{i {\hat p}\cdot({\hat x}-{\hat y})} \left\{
  \langle x_2|\psi_0\rangle \langle\psi_0|y_2\rangle \frac{-i \not \!
    {\hat p}}{{\hat p}^2}{\mathcal P}_R \right.
$$
$$
+\sum_{n=1}^\infty [\langle
x_2|\psi_n\rangle\langle\psi_n|y_2\rangle \frac{-i \not \! {\hat
    p}}{{\hat p}^2+\lambda_n^2}{\mathcal P}_R +\,\langle x_2|{\tilde
  \psi}_n\rangle \langle{\tilde \psi}_n|y_2\rangle \frac{-i \not \!
  {\hat p}}{{\hat p}^2+\lambda_n^2} {\mathcal P}_L
$$
\begin{equation}
  \label{eq:ferexp}
\left. +\,\sum_{n=1}^\infty [\langle x_2|{\tilde \psi}_n\rangle
\langle\psi_n|y_2\rangle \frac{\lambda_n}{{\hat p}^2+\lambda_n^2} 
{\mathcal P}_R +\,\langle x_2|\psi_n\rangle \langle{\tilde \psi}_n|y_2\rangle 
\frac{\lambda_n}{{\hat p}^2+\lambda_n^2} 
{\mathcal P}_L ] \right\} \;.
\end{equation}
We are using a discrete sum notation here for the sake of simplicity;
it should be noted, however, that with our assumptions on the potential,
there is only one discrete state, $\psi_0$. We shall take care of this,
using a more explicit notation when necessary.

On the other hand, for $\Delta_\varphi$ we use instead the expansion:
\begin{equation}
  \label{eq:bosexp}
\langle x|\Delta_\varphi|y\rangle \;=\; - \int \frac{d^2{\hat k}}{(2\pi)^2} \,
e^{i {\hat k}\cdot({\hat x}-{\hat y})}
\sum_{n=0}^\infty \langle x_2|\xi_n\rangle \langle\xi_n|y_2\rangle 
\frac{1}{{\hat k}^2 + \mu_n^2}\;.
\end{equation}
Of course, the same remark on the number of discrete modes
(now there are two of them), applies here.

In terms of the kernels defined above, we see that
\begin{equation}
 \label{eq:ddg1}
{\mathcal D}_1(x,y) \;=\; -
\frac{\delta^2 \Gamma_1}{\delta{\bar\chi}(x)\delta\chi(y)}\;=\; 
g^2 \Delta_\varphi(y,x) \, \langle x|{\mathcal D}^{-1}|y\rangle \;,  
\end{equation}
whose explicit form in the mixed Fourier representation becomes:
\begin{equation}
{\mathcal D}_1(x,y)\;=\;\int\frac{d^2{\hat k}}{(2\pi)^2} \,
e^{i {\hat k} \cdot ({\hat x} - {\hat y})} \gamma_k (x_2,y_2) 
\end{equation}
with
$$
\gamma_k (x_2,y_2) \;=\; - g^2 \, \left\{
 \sum_{n=0}^\infty \,
 \xi_n(x_2) \psi_0(x_2) \xi^\dagger_n(y_2) \psi^\dagger_0(y_2) \,
  J({\hat k};\mu_n,0) {\mathcal P}_R \right.
$$
$$
+\,\sum_{n=0,m=1}^\infty\,[\xi_n(x_2)\psi_m(x_2)\xi^\dagger_n(y_2) 
\psi^\dagger_m(y_2) \, J({\hat k};\mu_n,\lambda_m) {\mathcal P}_R ] 
$$
$$
+\,\sum_{n=0, m=1}^\infty \, \xi_n(x_2) {\tilde\psi}_m(x_2) \xi^\dagger_n(y_2) 
{\tilde\psi}^\dagger_m(y_2) \, J({\hat k};\mu_n,\lambda_m) {\mathcal P}_L  
$$
$$
+ \, \sum_{n=0, m=1}^\infty \, \xi_n(x_2) {\tilde\psi}_m(x_2) \xi^\dagger_n(y_2) 
\psi^\dagger_m(y_2) \, I({\hat k};\mu_n,\lambda_m) {\mathcal P}_R  
$$
\begin{equation}
  \label{eq:gammak}
\left. + \, \sum_{n=0, m=1}^\infty \, \xi_n(x_2) \psi_m(x_2) \xi^\dagger_n(y_2) 
{\tilde\psi}^\dagger_m(y_2) \, I({\hat k};\mu_n,\lambda_m) {\mathcal P}_L \right\} \;,
\end{equation}
where we used a discrete notation for the sum over eigenvalues, but
of course an integral over the continuous part of the spectrum is
implicitly assumed. More precisely, we only have one discrete zero 
mode ($\lambda_0 =0$) for the fermionic part, plus two discrete modes with
$\mu_0^2 = 0$ and $\mu_1^2 = 3 \kappa^2$ for the bosonic part of the sums. 

The functions $I$ and $J$ appearing in (\ref{eq:gammak}) may be
interpreted as $1+1$ dimensional loop integrals. They are explicitly
given by
\begin{equation}
  \label{eq:defi}
I({\hat k}; M_1, M_2)\;=\; \int \frac{d^2 {\hat p}}{(2 \pi)^2}\,
\frac{M_2}{[({\hat k}-{\hat p})^2 + M_1^2]({\hat p}^2 + M_2^2)} 
\end{equation}
and
\begin{equation}
  \label{eq:defj}
J({\hat k}; M_1, M_2)\;=\; \int \frac{d^2 {\hat p}}{(2 \pi)^2}\,
\frac{-i\not \! {\hat p}}{[({\hat k}-{\hat p})^2 + M_1^2]({\hat p}^2 + M_2^2)} \;.
\end{equation}
These (UV convergent) functions may be evaluated by using standard
methods, what yields
$$
I({\hat k}; M_1, M_2)\;=\;\frac{M_2}{4\pi \sqrt{({\hat k}^2 + 4M^2)(k^2+\Delta^2)}} \,
$$
\begin{equation}
\ln\left[\frac{{\hat k}^2+M_1^2+M_2^2+\sqrt{({\hat k}^2 + 4M^2)(k^2+\Delta^2)}}{{\hat k}^2
+M_1^2+M_2^2-\sqrt{({\hat k}^2 + 4M^2)(k^2+\Delta^2)}}\right]\;,
\end{equation}
and
\begin{equation}
J({\hat k};M_1,M_2)\;=\; - \frac{i \not\!{\hat k}}{4\pi k^2 (\alpha^+ - \alpha^-)} 
\left[\alpha^+\ln(\frac{\alpha^+-1}{\alpha^+})-\alpha^-\ln(\frac{\alpha^--1}{\alpha^-})\right]
\end{equation}
where 
\begin{equation}
\alpha^\pm\;=\; \frac{1}{2{\hat k}^2}
\left[{\hat k}^2 + 2 M \Delta \mp \sqrt{({\hat k}^2 + 4M^2)(k^2+\Delta^2)}\right] \;.
\end{equation}
and we used the notation: $M=\frac{M_1+M_2}{2}$, $\Delta=M_1-M_2$.

We now consider the effect of including this correction into the
Dirac operator. The corrected Dirac operator ${\tilde{\mathcal D}}$, 
equation (\ref{eq:fer2}), in a representation where the first two 
components have been Fourier transformed, will have a kernel 
\begin{equation} 
\langle x_2|{\tilde \mathcal D}|y_2\rangle = \langle x_2|(a 
+ i {\hat{\not \! k}})|y_2\rangle {\mathcal P}_L
+ \langle x_2|(a^\dagger + i {\hat{\not \! k}})|y_2\rangle 
{\mathcal P}_R + \gamma_K(x_2,y_2)
\end{equation}
where the `free' part is of course local and translation invariant,
since 
\begin{equation}
\langle x_2|a|y_2\rangle = a(x_2) \delta(x_2-y_2) 
\end{equation}
and the analogous relation for $a^\dagger$.
The correction $\gamma_k$, in turn, will be non local, with an structure
which may be described as follows:
$$
\gamma_k(x_2,y_2) \,=\, u_k(x_2,y_2) {\mathcal P}_L  + {\tilde u}_k(x_2,y_2) {\mathcal P}_R
$$
\begin{equation}
+  v_k(x_2,y_2) i \hat{\not\! k}{\mathcal P}_L  +
{\tilde v}_k(x_2,y_2) i\hat{\not\! k} {\mathcal P}_R \,,
\end{equation}
where the functions $u_k, {\tilde u_k}, v_k, {\tilde v_k}$ have a
rather complicated expression, as may be inferred from 
(\ref{eq:gammak}). 

To study the stability of the chiral zero mode solution, we note
that the zero mode $|\Psi_0\rangle$ of the `free' Dirac operator, ${\mathcal D}$, 
may be written as: $|\Psi_0\rangle = {\mathcal P}_L \psi_0(x_2) |k\rangle$, were $|k\rangle$ is a 
zero mass spinor, solution of the $1+1$ dimensional Dirac equation.
If we apply now the corrected Dirac operator $\tilde{\mathcal D}$ to
the same state $|\Psi_0\rangle$, we see that:
\begin{equation}
\langle x_2|{\tilde \mathcal D}|\Psi_0\rangle \,=\, 0 + \int_{-\infty}^{+\infty}dy_2\, \gamma_k(x_2,y_2) \langle y_2|\Psi_0\rangle \;,
\end{equation}
and let us now quantify the effect of the rhs. From the explicit form
of $\gamma_k$, we see that only the terms proportional to ${\mathcal P}_L$
may contribute. Moreover, the integrals over the $y_2$ coordinate will
be of the form:
\begin{equation}
\int_{-\infty}^{+\infty}dy_2\, \xi^\dagger_n(y_2) {\tilde \psi}^\dagger_m (y_2) \psi_0(y_2) \;=\; \xi_{nm} \;.
\end{equation}
The explicit forms of the functions appearing in the integral are
\begin{equation}
\psi_0(y_2) \,=\, \sqrt{\frac{\kappa}{2}} \, {\rm sech} (\kappa y_2) 
\end{equation}
for $\psi_0$, and 
\begin{equation}
{\tilde \psi}^\dagger_{m\neq0} (y_2) \,=\,\frac{1}{\sqrt{2\pi}} \, e^{i \sqrt{\lambda^2_m - \kappa^2} y_2} 
\end{equation}
since ${\tilde h}_f$ is a (shifted) free particle one dimensional 
Hamiltonian. Of course, $\lambda^2_n \geq \kappa^2$.
Regarding $\xi_n$, it has different profiles depending on whether
$n=0$, $n=1$ or belongs to the continuum:
\begin{equation}
\xi_0(y_2) \,=\, \sqrt{\frac{3 \kappa}{4}} \, {\rm sech}^2( \kappa y_2)\;,
\end{equation}
\begin{equation}
\xi^\dagger_1(y_2) \,=\, b^\dagger \psi_0(y_2) \,=\, N 
(-\partial_2 + 2 \kappa \tanh(\kappa y_2) ) \cosh(\kappa y_2) 
\,=\, \sqrt{\frac{3 \kappa}{2}} \, \frac{\sinh(\kappa y_2)}{\cosh^2(\kappa y_2)} \,
\end{equation}
and for states in the continumm, 
\begin{equation}
\xi^\dagger_n(y_2) \,=\,\frac{1}{\sqrt{2\pi}} \, 
(- i \sqrt{\mu^2_n - 4 \kappa^2}  + 2 \kappa \tanh( \kappa y_2) )\,
 e^{-i \sqrt{\mu^2_n - 4 \kappa^2} y_2} \;,
\end{equation}
where $\mu_n^2 \geq 4 \kappa^2$.

We now show  that $\xi_{n,m}$ is exponentially small. When $n=0$,
we have:
$$
\int_{-\infty}^{+\infty}dy_2\, \xi^\dagger_0(y_2) {\tilde \psi}^\dagger_m (y_2) \psi_0(y_2) \;=\; 
$$
$$
\frac{\kappa}{2} \, \sqrt{\frac{3}{\pi}}  \, 
\int_0^{+\infty} dy_2 \,\cos [\sqrt{\lambda^2_m - \kappa^2} y_2] {\rm sech}^3( \kappa y_2)
$$
\begin{equation}
=\,\frac{3 \sqrt{\pi}}{8} \, (\frac{\lambda_m}{\kappa})^2 \, 
{\rm sech}[\frac{\pi}{2} \sqrt{(\frac{\lambda_m}{\kappa})^2 -1}]  
\end{equation}
which is exponentially decreasing with $\lambda_m$.
For the $n=1$ case, we have instead:
$$
\int_{-\infty}^{+\infty}dy_2\, \xi^\dagger_1(y_2) {\tilde \psi}^\dagger_m (y_2) \psi_0(y_2) \;=\; 
$$
\begin{equation}
=\,- i \frac{1}{4} \sqrt{\frac{3\pi}{2}} \,  ((\frac{\lambda_m}{\kappa})^2 -1) \,
{\rm sech}[\frac{\pi}{2} \sqrt{(\frac{\lambda_m}{\kappa})^2 -1}]  
\end{equation}
which is also decreases exponentially.

When $n$ belongs to the continuum, a very similar inequality 
shows that $\xi_{n,m}$ is also exponentially small, and indeed much 
smaller than in the $n=0$ case.

All this shows that the free zero mode $|\Psi_0\rangle$ is still a zero mode of
the corrected Dirac operator, modulo exponentially small corrections.
It is noteworthy that the parameter that regulates the rate of decrease
of the exponentials is $\kappa$. Thus, the sharper the domain wall, the
more negligible the departures shall be. Of course, the fact that
the correction is so small suggests the use of the free zero mode as
a starting point for a perturbative evaluation of the corrected one,
by repeated application of the corrected Dirac operator.

It is interesting to see what happens when the gap between the zero modes
and the higher modes is very large, at the level of the interaction part 
of the effective action $\Gamma_1^{(I)}$, equation (\ref{eq:gmi}). 
When $\kappa$ becomes very large, we may  approximate the propagators
as follows:
$$
{\mathcal D}(x,y) \,\sim\, \psi_0 (x_2) \psi_0 (y_2)\, 
{\hat\mathcal D}({\hat x},{\hat y}) 
\;\;\;,\;\;
{\hat\mathcal D}({\hat x},{\hat y}) \,=\,
- \langle {\hat x}| {\hat{\not \! \partial}}^{-1} {\mathcal P}_R 
| {\hat y} \rangle \
$$
\begin{equation}
\Delta_\varphi(x,y) \,\sim \, \xi_0 (x_2) \xi_0(y_2) 
\, {\hat\Delta}_\varphi({\hat x},\hat{y}) \;\;\;,\;\;
 {\hat\Delta}_\varphi({\hat x},\hat{y}) 
\,=\, - \, \langle {\hat x}| {\hat \partial}^{-2} | {\hat y} \rangle \;.
\end{equation} 
Also, by a similar argument to the one used for the fermionic propagator,
we see that the `classical fields' $\chi$ appearing in (\ref{eq:gmi}) may
be replaced by the projections:
\begin{equation}
\chi (x) \sim \psi_0(x_2) {\hat \chi} ({\hat x})
\end{equation}
where ${\hat \chi} ({\hat x})$ is an arbitrary two dimensional spinor. 
Using these approximations, we find that
\begin{equation}
\Gamma_1^{(I)}\;\sim \; \frac{1}{2} {\rm Tr} \ln \left[ \delta(x-y) 
- 2 g^2 \sqrt{\frac{\kappa}{3}} \xi_0(x_2) \phi_0^2(x_2) \,
 G({\hat x},{\hat y}) \right]  
\end{equation}
where
\begin{equation}
G({\hat x},{\hat y}) \;=\; \int_{{\hat z}} 
{\hat\Delta}_\varphi({\hat x},\hat{z}) {\bar{\hat \chi}} ({\hat z})
 {\hat\mathcal D}({\hat z},{\hat y})  {\hat \chi} ({\hat y}) \;.
\end{equation}
On the other hand, from the relation between $\psi_0$ and $\xi_0$, we note
that 
\begin{equation}
\xi_0 (x_2) \, \psi_0^2 (x_2) \;=\; \sqrt{\frac{\kappa}{3}} 
\langle x_2 | \xi_0 \rangle \langle \xi_0 | y_2 \rangle \;,
\end{equation}  
and finally, this implies:
\begin{equation}
\Gamma_1^{(I)}\;\sim \; \frac{1}{2} {\rm Tr} \ln \left[ \delta({\hat x}-
{\hat y}) - 2 g^2 \sqrt{\frac{\kappa}{3}} \, G({\hat x},{\hat y}) \right]  
\end{equation}
where the $x_2$ dependence has disappeared.

\section{Induced electric field}\label{sec:el}
Let us discuss here the application of some general results~\cite{ffl}
regarding the induced electric field seen by the fermionic zero mode,
to show that, for this kind of configuration, and in this
approximation, that electric field vanishes on the average.

The effective electric field will only be non-zero when the defect is
accelerated, and this can only happen, in our situation, when we
include the fluctuations in the scalar field. We recall that, when the
fluctuations are small and the defect is rectilinear, the formula that
yields the effective electric field $E_{eff}$ for the fermionic zero
mode~\cite{ffl} in a given scalar field configuration is:
\begin{equation}\label{eq:effel1}
E_{eff} \;=\; \frac{1}{6} \left[\partial_0^2\eta \partial_1^2 \eta \,-\, (\partial_0\partial_1 \eta )^2 \right]
\end{equation}
where $\eta(x_0,x_1)$ is the fluctuation of $\varphi$ around its static kink
configuration with center in $x_2=0$. Expression (\ref{eq:effel1}) is
the value of the electric field for a given $\varphi$ configuration; it is
of course interesting to consider its quantum average $\langle E_{eff}\rangle$, the
average defined by the Euclidean action of the model. Before
considering the specific form of the average, we note that the 
{\em integral\/} of $E_{eff}$ necessarily vanishes, since
\begin{equation}\label{eq:effel2}
\int dx_0 dx_1 \, E_{eff} \;=\; \frac{1}{6} \int dx_0 dx_1 \, \left[\partial_0^2\eta \partial_1^2 \eta \,-\,
(\partial_0\partial_1 \eta )^2 \right]\;=\; 0
\end{equation}
by an integration by parts. This has consequences for the spacetime
average of $\langle E_{eff}\rangle$, but it leaves room for the existence of a
locally non-zero $\langle E_{eff}\rangle$.  The form of this local effective
electric field can be written, in our approximation, as follows:
\begin{equation}\label{eq:effel3}
\langle E_{eff}(x_0,x_1) \rangle  \;=\; \frac{1}{6} \left[
\frac{\partial^2}{\partial x_0^2}\frac{\partial^2}{\partial y_1^2} -
\frac{\partial^2}{\partial x_0 \partial x_1} \frac{\partial^2}{\partial y_0 \partial y_1}
\right] \Delta_\varphi(x_0,x_1;y_0, y_1)|_{y_\mu \to x_\mu} \;,
\end{equation}
where $\Delta_\varphi$ is the `fluctuation operator' defined in (\ref{eq:delphi}).
Now it is evident, since $\Delta_\varphi$ is translation invariant in $x_0$ and
$x_1$, that (\ref{eq:effel3}) vanishes. Indeed, translation invariance
implies that $\Delta_\varphi$ can only depend on the differences of its
arguments, and this makes it possible to show that the two terms on
the right hand side of  (\ref{eq:effel3}) cancel out.

This negative answer for the quantum average might of course be
expected on symmetry grounds. However, it can be used the other way
around, to look for the existence of a non-vanishing average by
considering different scalar field configurations
violating the invariance under translations in more than one coordinate.

\section{Conclusions}\label{sec:conc}
We have studied a system consisting of a Dirac fermion interacting 
with a real scalar field, to understand the effect of quantum 
fluctuations on the zero modes due to the Callan and
Harvey mechanism, to one-loop order.
We have shown that the effect of the fluctuations on the zero mode
is exponentially suppressed, and hence negligible, what signals the
perturbative stability of the classical zero mode solution under 
quantum fluctuations.

The effective inertial electric field induced on the fermions vanishes
to this order; and this also points to the direction of stability, since
an strong electric field on the zero mode may have meant that an
extra interaction should have been included.

Based on these results, we conclude that instabilities may very
likely arise from different configurations, possible with more than
one defect, and with translation invariance along only one spacetime
coordinate. Also, it remains an open question whether a nonperturbative
treatment for the quauntum coupled system is possible.  Nonperturbative
results do indeed exist for this system, although they deal with the 
external field problem only~\cite{DaRold:2001ri}.

Results on these systems, as well as on numerical nonperturbative 
solutions of the coupled system, will be reported elsewhere.
 
%
\section*{Aknowledgments}
The author acknowledges Prof.~I.~R.~Aitchison (Oxford) for many useful 
comments and suggestions.
Financial support from CONICET and Fundaci{\'o}n Antorchas (Argentina) is
also acknowledged.
%
\section*{Appendix A: Supersymmetric model}
We discuss here some particularities of the (simplest) supersymmetric 
version of the model discussed previously. Taking into account the 
fact that the bosonic part of the model has to be a real 
scalar~\footnote{This must be so in order to have domain wall like
solutions and fermionic zero modes.}, and we want a non-trivial 
interaction between boson and fermion fields, we are lead to consider 
a \mbox{$2+1$} dimensional Wess-Zumino model 
supplemented by a mass term and an interaction term. 

Following the notation and conventions of~\cite{susy}, we consider
the action 
\begin{equation}\label{eq:defssusy}
S\;=\; S_{WZ} \,+\, S_m \,+\, S_g
\end{equation} 
where $S_{WZ}$ is the free, massless, Wess-Zumino action
\begin{equation}\label{eq:defswz}
S_{WZ}\,=\, \int d^3x \left[ -\frac{1}{2} \partial^\mu \phi 
\partial_\mu \phi - \frac{1}{2} {\bar \psi} \not \!\partial \psi 
+ \frac{\alpha}{2} F^2 \right]
\end{equation}
where we used a Minkowski space metric \mbox{$\eta_{\mu\nu}={\rm diag}(-1,1,1)$},
and $\psi$ is a (real) Majorana fermion.  
The actions $S_m$ and $S_g$ are mass and interaction terms,
respectively:
$$
S_m\;=\; \int d^3x \; \frac{\kappa}{2} 
\left( F \phi - \frac{1}{2} {\bar\psi}\psi \right)
$$
\begin{equation}
S_g\;=\; \int d^3x\; g \left( F \phi^2 - {\bar\psi}\psi \phi \right) \;.
\end{equation}
As usual, the role of the auxiliary field $F$ is to make the action $S$ 
of-shell invariant under the supersymmetry transformations, which, as shown 
in~\cite{susy}, are given by: 
\begin{eqnarray}\label{eq:susytrans}
\delta \phi &=& {\bar\epsilon} \psi \nonumber\\
\delta F &=& {\bar\epsilon} \not \! \partial \psi  \nonumber\\
\delta\psi &=& \not\! \partial \phi \epsilon \,+\, F \epsilon \;.
\end{eqnarray}
Supersymmetry is preserved, although now on-shell, if the auxiliary field 
$F$ is `integrated out'. This yields the equivalent action
\begin{equation}
S\;=\; \int d^3x \, \left[ -\frac{1}{2} \partial^\mu \phi 
\partial_\mu \phi \,-\, \frac{1}{2} {\bar \psi} \not \!\partial \psi 
\,-\,\frac{1}{2} (\frac{\kappa}{2} \phi + g \phi^2)^2 \,-\, 
\frac{1}{2} (\frac{\kappa}{2} + 2 g \phi) {\bar\psi} \psi \right] \;.
\end{equation}

In order to compare with the non supersymmetric action, it is convenient to
make the field redefinition $\phi (x) \to \varphi(x)$, with 
\mbox{$\varphi (x) = \phi(x) + \frac{\kappa}{4g}$}  
so that, in the new variables, the action is:
\begin{equation}
S\;=\;\int d^3x \; \left[-\frac{1}{2} \partial_\mu \varphi \partial^\mu \varphi 
- \frac{1}{2} {\bar\psi} \not \! \partial \psi - \frac{g^2}{2} 
( \varphi^2 - (\frac{\kappa}{g})^2 )^2 
- g \varphi {\bar \psi} \psi \right] \;.
\end{equation}
This action is almost identical to its non-supersymmetric version, differing
only in their fermionic sectors. 


\end{document}